\documentclass{procmult}
\usepackage{graphicx}

\begin{document}

\title{Measuring solar disk shape up to relativistic accuracy: the role of scintillation in ancient naked eye data}
\author{Costantino Sigismondi$^\dag$, Richard Nugent$^\ddag$ \and Gerhard Dangl$^\S$}
\institute{
$^\dag$  ICRA, International Center for Relativistic Astrophysics, P.le Aldo Moro 5 00185 Roma, Italy, www.icra.it/solar, sigismondi@icra.it \\
$^\ddag$ IOTA, International Occultation Timing Association, US Section, rnugent@wt.net  \\
$^\S$    IOTA, European Section, www.dangl.at, gerhard@dangl.at }
\maketitle
\abstract{The orbital displacement of the Moon is known to a precision on the order of  centimeters from the lunar laser ranging data, while the lunar profile's confidence level is still only about 200 m. 
The lunar motion is used to measure the solar diameter during central eclipses with accurate timing of Baily's beads at the umbral path limits.
The onset or disappearance of a Baily's bead is due to the alignment between lunar limb's valleys and solar photosphere: this timing is determined outside our atmosphere. Due to scintillation, naked eye evaluations of such phenomena can be systematically different with respect to telescopes.
We analyze the case of  Venus' occultations as bead-like situation, in order to infer considerations on naked eye reports in historical total eclipse reports.

{\bf keywords:}
lunar occultation, scintillation, total eclipse, solar secular variability, solar diameter and oblateness, data analysis, scaling phenomena.
}

\section{Introduction}
During a lunar occultation the Moon covers a star or a planet. A solar eclipse is also, technically speaking, an occultation.
The role of ancient occultations, reported by Ptolemy in the Almagest, has been crucial to determine the famous secular acceleration of the lunar mean motion (Fotheringham, Litt and Longbottom, 1915; Fotheringham and Litt, 1923 and Mignard, 2004). Since the average angular lunar motion is about 0.5"/s, even with a rather low timing accuracy of about 10 minutes, the lunar motion across the sky covers 5 arcminutes. After more than 20 centuries those occultation where useful to detect the acceleration of $\sim-23"/cy^2$. 
More recently occultations have been used to make high resolution measurements on double stars, and Quasars exploiting the light curves (Leinert, 2004).

Asteroidal occultations allow to verify the gravitational light bending of the solar field at 50 degrees of elongation from it (Sigismondi and Troise, 2008) where it is only 6 mas (milliarcseconds), but this figure at 1 AU corresponds to 420 m, similarly lunar occultations occur in a different part of the solar field and the stars are shifted by a few mas with respect to their unperturbed position (Sigismondi, 2007 submitted to JKPS).
Lunar laser ranging deals with the position of the optical center of the Moon, averaged upon several years of measurements. The accuracy to several millimeters, achieved after more than three decades of measurements, allows us to determine relativistic phenomena such as geodetic or de Sitter precession of the orbit.

Finally, special types of occultation as total solar eclipses, are used to recover a milliarcsecond accuracy in solar disk shape measurements, up to relativistic interests dealing with the solar oblateness. This is done from the timing Baily's beads.
Among those data we have some historical reports made with naked eye (Sigismondi, Bianda and Arnaud, 2008 on 1715 and 1869 total eclipses). Since those observations have been made near the edges of totality they are very useful to recover the ancient diameter of the Sun, and here we want to study the role of scintillation.
In each eclipse we consider as unknown the correction to the mean angular solar radius of 959.63", but in order to assess the influence of scintillation it is better to have an occultation of Venus whose diameter is known.
We have examined an observation made with 60 fps camcorder with a 21 mm objective in parallel with naked eye on Dec. 1st, 2008 during the last Venus' occultation visible in Europe for several years to come (Herald, 2008).

\section{Data}

The observation of December 1, 2008 was done at the Pontifical University Regina Apostolorum, UPRA, in Rome, 
with a binocular TASCO 8x21 mm in front of a SANYO CG9 videocamera the occultation of December 1st, 2008 has been recorded at 60 fps rate on 640x480 pixel, 800 ISO, H.264 format. The original video is MPEG-4 and the detector is CMOS.
The Moon was about 164 pixels  diameter while Venus saturated about 4 pixel; their angular diameters were about 1800" vs 17" in a ratio 100:1.
The phase of Venus was 0.69.

\begin{figure}
 \centering
	\includegraphics{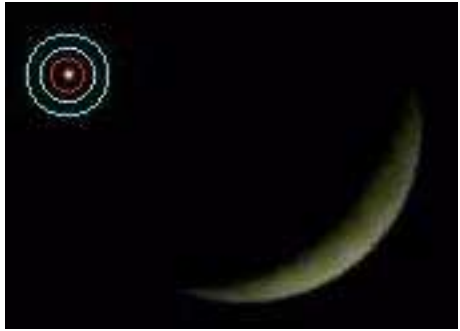}
\caption{Venus and the Moon; Venus is with three circles superimposed: the inner circle is the signal's integration area, while in the outer ring's area the background is integrated.}
\label{figure}
\end{figure}

\begin{figure}
 \centering
	\includegraphics{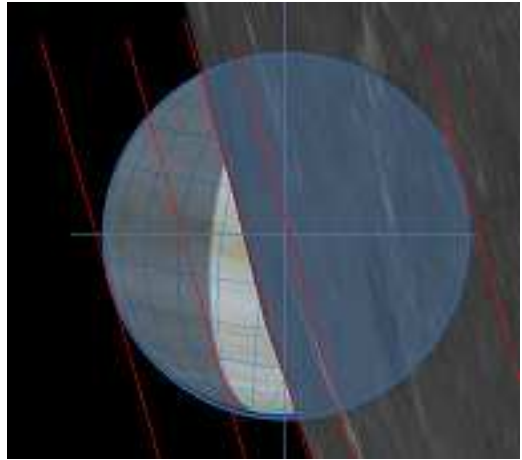}
\caption{Geometrical circumstances of the dec. 1st, 2008 occultation of Venus. The image is simulated at 16:18:21 s UTC of december 1st, 2008 for observer's location 12.3946E 41.8795N 66 m above sea level with Guide8 software. The last glimpse of venusian light comes from a decreasing circular segment. Parametrizing its decrease with time, the exposed area is $\propto (t_o-t)^{3/2}$ where $t_o \le t$ is the disappearing time.}
\label{figure}
\end{figure}

\section{Analysis}

By visual inspection frame per frame, with QuickTime 7.0 the last glimpse of Venus is at 68.40 s, video time, while
playing the video, which contained also an audio record, it has been possibile to find when people started to say that Venus was disappeared, i.e. between 62 and 63 s, video time.
The last glimpse is preceded by several dark frames, due to the strong scintillation effect.
An analysis with Limovie software (Miyashita, 2008; Nugent, 2008) has shown a more detailed structure.
The last glimpse as detected by Limovie is at 66.90 s, video time. 
The difference with naked eye inspection is probably due to some background pixel's flash.
Fitting the signal with a $(t_o-t)^{3/2}$ curve leads to $t_o=$65.84 s, lower than the last glimpse.
Modifying the signal S with an exponential factor exp(S/a), with a=800...20000, the final fitting yields $t_o=$64.84 s...65.90 s. This is an indication that the SANYO videocamera's response is not linear, and this is the reason why $t_o$ from $(t_o-t)^{3/2}$ fit is extrapolated before the last light signal left in the video.

\begin{figure}
 \centering
	\includegraphics{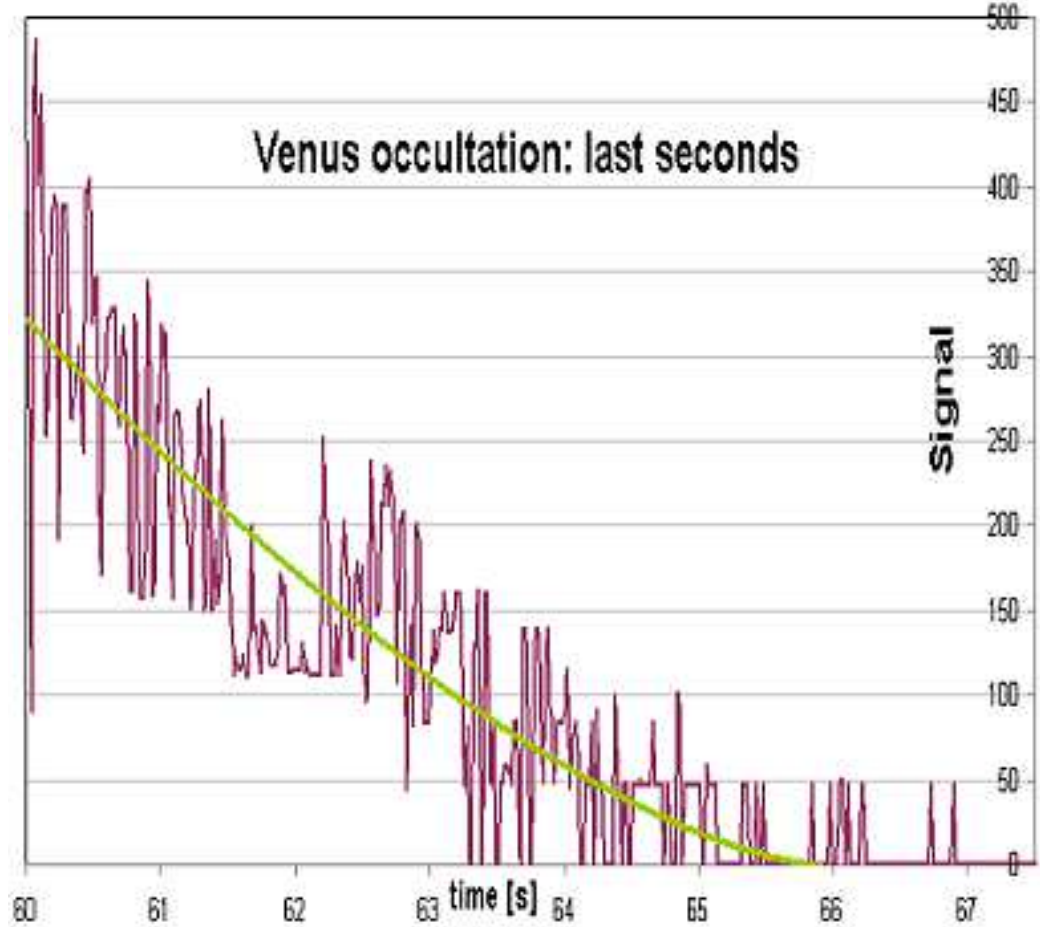}
\caption{Plot of the signal from final seconds of the Venus' occultation, as detected by Limovie. A geometrical fit $\propto (t_o-t)^{3/2}$ is superimposed. UTC timing is video time + 12.23 s + 16 h 20 m.}
\label{figure}
\end{figure}

\section{Discussion}
The luminosity of Venus, with visual magnitude of -4.7, corresponds to a surface luminosity 10 magnitudes lower than the photosphere, so its disappearance can be considered as the study of a Baily's bead 10000 times smaller in area, 100 in dimension.
All conclusions in timing studies have to be rescaled by a factor of 100 to export these results in total eclipses, with similar geometry, where the incoming light from the photosphere is able to saturate immediately all unfiltered camcorders in the first second of appearance. Such scaling, in reality, has to face the Limb Darkening Function of the Sun, for which the last arcsecond of photosphere is rapidly fading to 16\% of the center-disk luminosity level.

The difference of $4\div5$ seconds in the eye versus instrumental detection would become 0.05 s during a total eclipse, with similar background luminosity and occulting geometry. Near the shadow's edge, where the beads are nearly grazing, the dis/appearance of the beads is a more long process and scintillation becomes important.
The reliability of naked eye observations for solar diameter historical measurements is therefore proved only in centerline. Obviously the relative timing has to be fairly efficient, and it has to be controlled on the historical sources available.

\section{Conclusions}

The powerful analysis tool of Limovie is presented along with the first 60 fps portable detector. Its suitability for astronomical observations is discussed: even if its response is not linear, because of automatizations, in the next total eclipse this will be employed to record Baily's beads with twice the accuracy of previous video. 
The role of scintillation during total eclipse is strictly related with flying shadows: when the eclipsed Sun is point-like as it undergoes such phenomenon with a surface luminosity 10000 larger than Venus.
By studying the Venus' occultation and scaling for luminosity and dimensions to the photospheric situation we have demonstrated that naked eye can be $0.04\div0.05$ s in advance with respect to instruments to individuate the disappearance of a bead. 
The eclipse video time resolution depends on frame rate: a millisecond accuracy is reliable from this scaling process.

By avoiding cases of bleaching eyes, easily occurring during eclipse viewing, the eye is as trustable as an instrument. All problems concern the method of timely recording the event. 

Finally at the Moon's distance the Fresnel scale is 10 m, which at lunar orbital speed of 1 Km/s is 0.01 s. To have Fresnel fringes it is necessary that the source is pointlike, i.e. smaller than 5 mas, the angle of 10 m at mean lunar distance.
This is not the case neither for Venus nor for the Sun.
The relativistic issue of measuring solar oblateness requires milliarcsecond accuracy, or a millisecond accuracy in timing (Herchak, 2007). Taking the last glimpse of light avoids problems with fitting scintillation, during grazing beads (polar), while both solutions, fitting or last glimpse, work with similar efficiency in equatorial beads.
  


\begin{thebibliography}{99}




\bibitem{Toomer} Toomer, G. J., {\it Ptolemy's Almagest} (Princeton University Press, 1998)

\bibitem{FL15}
J. K. Fotheringham, D. Litt,  and G. Longbottom, {\it MNRAS} {\bf 75} 377 (1915).

\bibitem{FL23}
J. K. Fotheringham, D. Litt {\it MNRAS} {\bf 83} 370 (1915).

\bibitem{M04} F. Mignard, 
{\verb|http://www.oca.eu/Mignard/Grex/Presentations_pdf/Grex04_F_Mignard.pdf|}2004

\bibitem{Leinert} Ch. Leinert, et al. 
{\verb|http://www.caha.es/newsletter/news02a/leinert/leinert.pdf|}2004

\bibitem{SigiTroise08}
C. Sigismondi and D. Troise, {\it Proc. XI Marcel Grossmann Meeting, R. Ruffini, R. T. Jantzen and H. Kleinert eds.} 2594 (World Scientific Pub. Singapore, 2008)

\bibitem{SigiBiandArnaud}
C. Sigismondi, M. Bianda and J. Arnaud, {\it (AIP Conf. Proc.} {\bf 1059} 189 (2008).

\bibitem{Herald}  D. Herald, {\it Canon of Lunar Occultations of Venus 2000 to 2100} private comm.(2009). 


\bibitem{limovie}
K. Miyashita,{\verb|http://www005.upp.so-net.ne.jp/k_miyash/occ02/limovie.html|} 2008 


\bibitem{RN08}
R. Nugent ed., Chasing the Shadow. The IOTA Occultation Observer's Manual 
{\verb|http://www.poyntsource.com/IOTAmanual/index.htm|} (IOTA, 2008)

\bibitem{milli}
S. Herchak, 
{\verb|http://www.lunar-occultations.com/iota/vti2.pdf|}(2007). 

\bibitem{G8}
{\verb| http://www.projectpluto.com/guide8.htm|} (2008).

\bibitem{H}
Halley, E., {\it Observation of Late Total Solar Eclipse on
the 22nd, April [...]}, {\it Phil. Trans.
R. Soc. of London} {\bf XXIX} 245-262 (1714-1716).

\bibitem{New}
Newcomb, S., in {\it Reports on observations of the eclipse
of august 7, 1869}, p. 15-22, U.S. Naval Observatory
(1870).

\bibitem{Rog}
Rogerson, B.D., {\it The Solar Limb Intensity Profile},
{\it Astrophysical Journal} {\bf 130}, 985 (1959).


\end{thebibliography}
\end{document}